\documentclass[12pt]{iopart}

\begin{document}

\title{ Comparison between entangled and nonentangled two-reservoir Kondo systems}

\author{Jongbae Hong}

\address{Department of Physics, POSTECH and
Asia Pacific Center for Theoretical Physics, Pohang, Gyeongbuk 790-784, Korea }
\ead{jbhong@postech.ac.kr}
\begin{abstract}

We clarify the difference between entangled and nonentangled two-reservoir mesoscopic Kondo systems and reveal the reason why theories using the Keldysh formalism, quantum Monte Carlo calculations, and
the renormalization group approaches cannot explain the line shapes of tunneling conductance of mesoscopic Kondo systems measured by using a two-terminal setup but explain those of a three-terminal setup. We emphasize that the previous theories study a nonentangled system, while real two-reservoir mesoscopic Kondo systems are entangled systems in which two reservoirs are within the coherent region. We show that two coherent side peaks appearing in tunneling conductance signify the entanglement between two reservoirs. These side peaks are essential for explaining the experimental observations for tunneling conductance.

\end{abstract}



\maketitle

\section{Introduction}
Coherent transport in a mesoscopic Kondo system has been attracting wide interest since the discovery of
the Kondo effect in a quantum dot single-electron transistor (QDSET)~\cite{gordon}. Observations of tunneling
conductance for the QDSET~\cite{wiel} and other Kondo-involved mesoscopic systems, such as a quantum point
contact (QPC)~\cite{cronen,qpc} and magnetized atom adsorbed on an insulating layer covering a metallic
substrate~\cite{otte,choi}, have been reported. These tunneling conductances demonstrate novel Kondo
phenomena observed at steady-state nonequilibrium. The two-reservoir Anderson impurity model under bias is
considered a proper microscopic model for describing the above-mentioned mesoscopic Kondo systems.
However, the nonlinear line shapes of tunneling conductance of those systems are not clearly explained
theoretically. Previous theoretical studies using the real-time renormalization group (RG)
method~\cite{schoeller}, the Keldysh formalism~\cite{wingreen2,fujii},
quantum Monte Carlo calculations~\cite{han}, and the scattering-state numerical RG method~\cite{anders} cannot explain the line shapes of the tunneling conductance of the above-mentioned systems,
especially the two side peaks shown in the QPC and adsorbed magnetized atom.

\begin{figure}
[b] \vspace*{6.4cm} \includegraphics{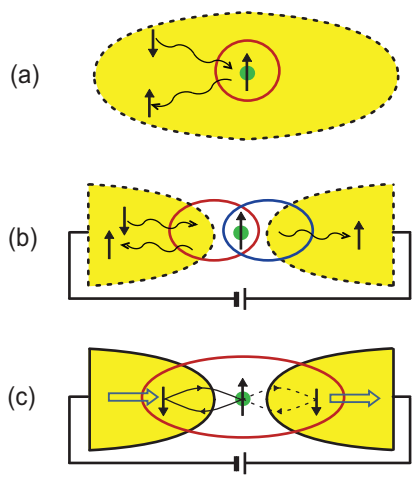} \vspace*{0.0cm}
\caption{(Color online) (a) Spin flip scattering in a conventional Kondo system. The red circle
indicates the coherent region. (b) The Hershfield model having large reservoirs.
The different color of the loop means nonentanglement.
(c) Realistic mesoscopic Kondo system with a single coherent region. Kondo singlets are entangled and the reservoir region inside the coherent region is a Kondo cloud. The open arrows indicate unidirectional motion of entangled Kondo singlet. }
\end{figure}

An unanswered question is then, ``Why are these approaches unable to reproduce the nonlinear $dI/dV$ line shapes,
where $I$ and $V$ denote current and source-drain bias voltage, respectively?"
Some of those theories are quite rigorous and sophisticated. Nevertheless, all of them commonly give a bias-dependent split Kondo peak in their spectral functions. In this study, we find the reasons for this from the schematic model and the methods used, and we provide an appropriate model and method to explain the experimental tunneling conductances measured by using two-terminal setup.

We first discuss the schematic employed in the previous approaches~\cite{schoeller,wingreen2,fujii,han,anders} and then discuss their methods. The previous approaches use the schematic suggested by Hershfield~\cite{hersh1} who adopted infinitely large reservoirs and considered spin scattering by a Kondo impurity, which was first studied by Nozi\`{e}res~\cite{nozi}. We depict spin-flip scattering, for example, in the conventional single-reservoir Kondo system in Fig.~1(a) and in the Hershfield model for a two-reservoir Kondo system under bias in Fig.~1(b). The latter model is a simple extension of Fig.~1(a) for two reservoirs.
However, we believe that the coherent region in a realistic two-reservoir mesoscopic Kondo system covers parts of both metallic reservoirs as well as the mediating Kondo atom, as shown in Fig.~1(c). A big difference exists between the two models: The left and right Kondo singlets in Fig.~1(b) are not entangled, whereas those in Fig.~1(c) are entangled. In other words, the inter-reservoir coherence is broken in Fig.~1(b), whereas it is retained in Fig.~1(c).

\section{Entangled vs. Nonentangled}
The wave function in the region including the left $(L)$ and right $(R)$ Kondo clouds in Fig.~1(b) can be written in a separate form,  $|\Psi\rangle_{KS}^L=(1/2)\left(\mid\downarrow\uparrow\rangle_{KS}
+\mid\uparrow\downarrow\rangle_{KS}\right)_{L}$ and $|\Psi\rangle_{KS}^R=(1/2)\left(\mid\downarrow\uparrow\rangle_{KS}
+\mid\uparrow\downarrow\rangle_{KS}\right)_R$, where the subscript "$KS$" denotes Kondo singlet. These two wave functions are matched at the mediating Kondo atom to study the scattering process. The scattering-state NRG~\cite{anders} corresponds to this situation.
For the entangled Kondo singlet (EKS) depicted in Fig.~1(c), the wave function may be written as $|\Psi\rangle_{ent}=(1/2)(e^{i\phi_L}|\Psi\rangle_{KS}^L+e^{i\phi_R}|\Psi\rangle_{KS}^R)$ with phase parameters $\phi_L$ and $\phi_R$.
This type of wave function has been used by Feynman~\cite{feynman} to study the Josephson junction~\cite{joseph}. He showed that the phase difference induces a coherent supercurrent at zero bias (known as the dc Josephson effect).
Even though the two-level model of Feynman is too simple to apply to the EKS tunneling in the mesoscopic Kondo system, it is clear that the phase difference between two Kondo clouds in the mesoscopic Kondo system may play a similar role as it does in the Josephson junction.
In the following, we show a characteristic phenomenon of entangled system in which the Kondo peak of the EKS state at equilibrium has an extra spectral weight owing to inter-reservoir coherence in addition to the spectral weight of the non-EKS state, and this extra contribution suddenly vanishes when a bias is applied. One expects that a phenomenon similar to the dc Josephson effect could be observed in an entangled Kondo system if contact resistance is removed. Such a system is a mesoscopic metal ring with a gap in which a Kondo impurity is located. As a concluding remark, the essence in Fig.~1(c) is phase difference reflecting inter-reservoir coherence as far as wave function is concerned.

The previous studies~\cite{schoeller,wingreen2,fujii,han,anders} are based on the Keldysh formalism~\cite{keldysh} and the Hershfield density matrix. However, the Keldysh formalism suffers from a fundamental difficulty because the phase factors are not explicitly developed on the Keldysh contour, and it requires a perturbation scheme based on the state in the remote past at which the system does not have inter-reservoir coherence. In the real-time RG calculation~\cite{schoeller} avoiding perturbation, one traces out the degrees of freedom of the reservoirs. This process also breaks inter-reservoir coherence. Also, the Hershfield density matrix does not reach the ensemble of EKS states by a time-evolution operation from the remote past. Instead, one reaches a non-EKS state on the Keldysh contour and the time-evolution brings to the ensemble of non-EKS states, in which inter-reservoir coherence is broken.
In summary, the various methods used in the previous studies are consistent within the Hershfield model of Fig.~1(b), and they are valid for studying the tunneling of non-EKS. As a result, the previous studies produce similar types of spectral functions exhibiting the bias-dependent split Kondo peak observed in a three-terminal experiment~\cite{letur}, in which the probing terminal plays
the role of phase-breaking scatterer and detects the incoherent current~\cite{datta}.

\section{Method for entangled system}
The real mesoscopic Kondo system depicted in Fig. 1(c) requires study of EKS tunneling to explain the $dI/dV$ line shapes measured by using a two-terminal setup. We emphasize the effect of inter-reservoir coherence in understanding the experimental data correctly. The problem is how to retain the inter-reservoir coherence in a theoretical analysis or to identify what kind of theoretical formalism it makes possible.
One possible way is to obtain the Green's function of the resolvent operator form,
\begin{equation}\label{eq1}
G^{+}_{mm\uparrow}(\omega)=\langle c_{m\uparrow}|(\omega{\bf I}-{\rm\bf L})^{-1}|c_{m\uparrow}\rangle ,
\end{equation}
in terms of a complete set of basis vectors of the two-reservoir Anderson impurity model,
${\cal H}={\cal H}_0^L+{\cal
H}_0^R+\sum_{\sigma}\epsilon_mc^\dagger_{m\sigma} c_{m\sigma}
+Un_{m\uparrow}n_{m\downarrow}+{\cal H}_C,$
where ${\cal H}_0^{L,R}=\sum_{k,\sigma}(\epsilon_k-\mu^{L,R})c^{\dagger}_{k\sigma}
c_{k\sigma}$, ${\cal H}_C=\sum_{k,\sigma,\nu=L,R}(V_{km}^{\nu}c^\dagger
_{m\sigma}c_{k\sigma}+V^{\nu*}_{km}c^{\dagger}_{k\sigma}
c_{m\sigma})$, and $\sigma$, $\epsilon_{k}$, $\epsilon_{m}$,
$V_{km}$, $U$, and $\mu$ indicate the electron spin, kinetic
energy, energy level of the mediating atom, hybridization
strength, on-site Coulomb repulsion, and chemical potential,
respectively. In equation (\ref{eq1}), ${\bf I}$ and ${\rm\bf L}$ are the identity
and Liouville operators, respectively. The Liouville operator is defined by
${\rm\bf L}\cal O\equiv{\cal H}\cal O-\cal O\cal H$ for an operator $\cal O$
and the inner product is defined by
$\langle \cal A|\cal B\rangle\equiv\langle \cal A\cal B^\dagger+\cal B^\dagger \cal A\rangle$,
where $\cal B^\dagger$ is the adjoint of $\cal B$ and the angular brackets denote the expectation value.
We perform only the operator calculations to represent the inner products.
We do not obtain the expectation values and leave them as free parameters, thus avoiding
the difficulty in determining a correct nonequilibrium density matrix.
Using a complete set of basis vectors guarantees a description of the inter-reservoir coherence.

A complete set of basis vectors spanning the Liouville space of the two-reservoir Anderson
impurity model has been obtained in our previous study~\cite{hong11,hong10}.
Those basis vectors describing up-spin dynamics are divided into three groups: \\ I:
$\{c_{m\uparrow}, \, \, \, \, n_{m\downarrow}c_{m\uparrow},
 \, \, \, \, j^{\pm L}_{m\downarrow}c_{m\uparrow},  \, \, \, \, j^{\pm R}_{m\downarrow}c_{m\uparrow}\}$,\\
II: $\{({\rm\bf L}_C^nj^{\pm L}_{m\downarrow})c_{m\uparrow},  \, \, \, \,
({\rm\bf L}_C^nj^{\pm R}_{m\downarrow})c_{m\uparrow} \, |
 \, \, n=1, \dots, \infty\}$, and\\
III: $\{c_{k\uparrow}^{L,R}, n_{m\downarrow}c_{k\uparrow}^{L,R},
({\rm\bf L}_C^nj^{\pm L,R}_{m\downarrow})c^{L,R}_{k\uparrow}| n, k=0, 1, \cdots, \infty \}$,
where ${\rm\bf L}_C$ denotes the Liouville operator using ${\cal H}_C$,
$j^+_{m\downarrow}=\sum_k(V_{km}c^\dagger_{m\downarrow}c_{k\downarrow}+V^*_{km}c^\dagger_{k\downarrow}c_{m\downarrow})$,
and $j^-_{m\downarrow}=i\sum_k(V_{km}c^\dagger_{m\downarrow}c_{k\downarrow}-
V^*_{km}c^\dagger_{k\downarrow}c_{m\downarrow})$. The basis vectors in groups I and II
describe the mediating Kondo atom, whereas those in group III describe the reservoirs, and they are
used to represent self-energy.

We construct a working Liouville space by eliminating unimportant basis vectors.
We neglect multiple spin-exchanging processes because they rarely occur in the Kondo regime~\cite{hong10}
and are less likely to occur under bias. Hence, we eliminate group II. We construct self-energy using the
virtual processes only and the basis vectors $({\rm\bf L}_C^nj^{\pm L,R}_{m\downarrow})c^{L,R}_{k\uparrow}$
in group III are neglected. We further neglect the basis vector $n_{m\downarrow}c_{m\uparrow}$ in group I
because it considers double occupancy up to $U^\infty$ order by $n_{m\downarrow}^\infty=n_{m\downarrow}$ and we
study the large-$U$ regime. It is noteworthy that this reduction in degrees of freedom does not affect
retention of inter-reservoir coherence.

After reduction of the number of degrees of freedom, the working Liouville space is
spanned by
$$\{c_{k\uparrow}^L, \, \, \delta
n_{m\downarrow}c_{k\uparrow}^L, \, \,
\delta j^{+L}_{m\downarrow}c_{m\uparrow}, \, \, \delta
j^{-L}_{m\downarrow}c_{m\uparrow}, \, \, c_{m\uparrow}, \, \,
\delta j^{-R}_{m\downarrow}c_{m\uparrow}, \, \, \delta
j^{+R}_{m\downarrow}c_{m\uparrow}, \, \, \delta
n_{m\downarrow}c_{k\uparrow}^R, \, \,  c_{k\uparrow}^R\},$$ where
$k=0, 1, \cdots, \infty$ represent the quantum states of the reservoirs.
We use $\delta$ indicating $\delta A=A-\langle
A\rangle$ to achieve orthogonality among the basis vectors.
For convenience, we omit the normalization factors $\langle(\delta
j^{\pm L,R}_{m\downarrow})^2\rangle^{1/2}$ and $\langle(\delta
n_{m\downarrow})^2\rangle^{1/2}$ in the denominators of the
corresponding basis vectors.

In the schematic shown in Fig.~1(c), the EKS tunnels unidirectionally and the electron always passes through the mediating Kondo atom.
Therefore, the local density of states (LDOS) at the mediating site ``$m$", i.e., the spectral function
$\rho_{m\uparrow}^{ss}(\omega)=-(1/\pi){\rm Im}G^{+}_{mm\uparrow}(\omega)$, contains fundamental information.
In equation (\ref{eq1}), the LDOS is written as $\rho_{m\uparrow}^{ss}(\omega)=(1/\pi){\rm Re}[({\bf M})^{-1}]_{mm}$,
where the matrix elements of ${\bf M}$ are given by ${\rm\bf M}_{pq}=-i\omega\delta_{pq}+\langle{\hat e}_q|i{\rm\bf L}{\hat
e}_p\rangle$. Arranging the basis vectors of the working Liouville space in the order written above allows
the matrix ${\bf M}$ to be written as
\begin{equation}\label{eq4}
{\rm\bf M}=\left(
\begin{array}{ccc} {\rm\bf M}_{LL} & {\rm\bf M}_{CL} &
{\rm\bf 0} \\ {\rm\bf M}_{LC} &
{\rm\bf M}_{C} & {\rm\bf M}_{RC}
\\ {\rm\bf 0} & {\rm\bf M}_{CR}
& {\rm\bf M}_{RR}
\end{array} \right).
\end{equation}
The block ${\rm\bf M}_{LL}$ (${\rm\bf M}_{RR}$) is composed of two
infinite-dimensional diagonal blocks with elements $-i(\omega-\epsilon_k)$
that are constructed by the basis vectors $c_{k\uparrow}^{L(R)}$ and $\delta
n_{m\downarrow}c_{k\uparrow}^{L(R)}$ describing the left (right) reservoir;
${\rm\bf M}_{C}$ is a $5\times 5$ block constructed of five basis
vectors at the center describing the mediating Kondo atom. In contrast, ${\rm\bf M}_{CL}=
-{\rm\bf M}_{LC}^\dagger$ and ${\rm\bf M}_{CR}=-{\rm\bf
M}_{RC}^\dagger$ are $5\times\infty$ and $\infty\times 5$ blocks,
respectively. The block ${\rm\bf M}_{CL}$ is written as
\begin{equation}\label{eq5}
{\rm\bf M}_{CL}=\left[ \begin{array}{c c c c c} {\rm\bf 0} &
{\rm\bf 0} & {\rm\bf C}^L_{{\bf k}m} & {\rm\bf 0} & {\rm\bf 0} \\
{\rm\bf C}_{{\bf k}j^-}^{\, LL} & {\rm\bf C}_{{\bf k}j^+}^{\, LL} & {\rm\bf 0} &
{\rm\bf C}_{{\bf k}j^+}^{\, LR} & {\rm\bf C}_{{\bf k}j^-}^{\, LR}
\end{array} \right],
\end{equation}
where ${\rm\bf C}^L_{{\bf k}m}$, ${\rm\bf C}_{{\bf k}j^-}^{LL}$, and ${\rm\bf C}_{{\bf k}j^+}^{LR}$
are infinite-dimensional column vectors having elements
$iV^L_{km}$, $V^L_{km}\xi^L_{-}$, and $V^L_{km}\xi^R_+$, $k=0,\cdots,\infty$,
respectively, and
\begin{eqnarray}
&\xi^{L,R}_{\pm}=(1/2)[\langle i[n_{m\downarrow},j^{\pm L,R}_{m\downarrow}]
(1-2n_{m\uparrow})\rangle+i(1-2\langle
n_{m\downarrow}\rangle)\langle j^{\pm L,R}_{m\downarrow}\rangle]
\times& \nonumber \\&[\langle (\delta j^{\pm L,R}_{m\downarrow})^2\rangle\langle
(\delta n_{m\downarrow})^2\rangle]^{-1/2}.&\nonumber
\end{eqnarray}
${\rm\bf M}_{CR}$ is point symmetric for the center of ${\rm\bf M}_{CL}$.
All the blocks around ${\rm\bf M}_C$ are transformed to the self-energy matrix below and the central
block ${\rm\bf M}_C$  is given by
\begin{eqnarray}\label{eq6}
{\rm\bf M}_C=\left( \begin{array}{c l l c c} -i\omega' & \gamma_{_{LL}} &
-U^L_{j^-} & \gamma_{_{LR}} & \gamma_{_j} \\ -\gamma_{_{LL}} & -i\omega'
& -U^L_{j^+} & \gamma_{_j} & \gamma_{_{LR}} \\
U_{j^-}^{L*} &  U_{j^+}^{L*} & -i\omega' &  U^{R*}_{j^+} &
U^{R*}_{j^-} \\  -\gamma_{_{LR}} & -\gamma_{_j} & -U_{j^+}^R  &
-i\omega' & -\gamma_{_{RR}} \\
 -\gamma_{_j} &  -\gamma_{_{LR}} &
 -U_{j^-}^R  & \gamma_{_{RR}} &  -i\omega'
\end{array} \right),
\end{eqnarray}
where $\omega'\equiv\omega-\epsilon_m-U\langle n_{m\downarrow}\rangle$
and $\langle n_{m\downarrow}\rangle$ denotes the average number of
down-spin electrons occupying the mediating atom.

\section{Inter-reservoir coherence}
The inter-reservoir coherence is contained in
the matrix elements representing the left-right overlap, such as
${\rm\bf C}_{{\bf k}j^\mp}^{\, LR}$ in ${\rm\bf M}_{CL}$ and $\gamma_{_{LR}}$ and $\gamma_{_{j}}$
in ${\rm\bf M}_{C}$. The latter two are written as
$\gamma_{_{LR}}=\langle \widehat{V}\sum_{r\in R}\sum_{l\in
L}|V|^2(c^\dagger_{l\downarrow}c_{r\downarrow}+c^\dagger_{r\downarrow}c_{l\downarrow})\rangle$ and
$\gamma_{_{j}}=\langle \widehat{V}\sum_{r\in R}\sum_{l\in L}|V|^2
(c^\dagger_{l\downarrow}c_{r\downarrow}-c^\dagger_{r\downarrow}c_{l\downarrow})\rangle$, where
$\widehat{V}=\sum_kiV(c_{k\uparrow}^L+c_{k\uparrow}^R)c^\dagger_{m\uparrow}$~\cite{hong11}. We omit the normalization factors and set $V_{km}^{L,R}=V$ in $\widehat{V}$. Note that $|V|^2$ indicates double hopping to go from one reservoir to another and that $\gamma_{_{LR}}$ and $\gamma_{_{j}}$ represent inter-reservoir coherence. Figure~2  shows the operator dynamics of $\gamma_{_{LR}}$ and $\gamma_{_{j}}$ depicting a singlet hopping passing through the mediating site. The negative sign in the middle of Fig.~2 indicates that $\gamma_{_{j}}$ represents the effect of bias and vanishes at equilibrium. In contrast, the unidirectional motion of EKS tunneling under bias requires the equality $\gamma_{_{j}}=\gamma_{_{LR}}$, which is the condition of steady-state nonequilibrium.
However, $\gamma_{_{LL(RR)}}=\langle\widehat{V}[j^{-L(R)}_{m\downarrow},j^{+L(R)}_{m\downarrow}]\rangle$
represents the degree of Kondo coupling on the left (right) side of the mediating Kondo atom
and $U_{j^\pm}^{L,R}=U\langle(\delta n_{m\downarrow})^2\rangle^{1/2}\xi^{L,R}_{\pm}$ represents
the degrees of double occupancy coming from the left or right reservoir via the incoherent motion represented by
$j^{-}_{m\downarrow}$ or $j^{+}_{m\downarrow}$.

\begin{figure}
[b] \vspace*{2cm} \includegraphics{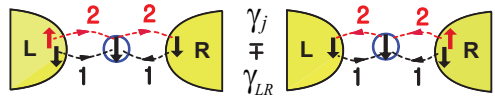} \vspace*{0.0cm} \caption{ (Color
online) Spin dynamics in $\gamma_{_{j}}$ (minus sign) and $\gamma_{_{LR}}$ (plus sign).
In equilibrium, $\gamma_{_{j}}=0$ and under bias, $\gamma_{_{j}}=\gamma_{_{LR}}$.
The numbers denote the sequence of coherent motion performing singlet hopping. }
\end{figure}

The infinite-dimensional matrix ${\rm\bf M}$ in equation (\ref{eq4}) can be transformed into a finite-dimensional matrix via matrix reduction~\cite{loewdin}, giving ${\rm\bf M}^{r}={\rm\bf M}_{C}-{\rm\bf
M}_{LC}{\rm\bf M}_{LL}^{-1}{\rm\bf M}_{CL}-{\rm\bf M}_{RC}{\rm\bf
M}_{RR}^{-1}{\rm\bf M}_{CR}$~\cite{hong11}. The last two terms form the self-energy matrix whose elements are given by $i\Sigma_{pq}=\beta_{pq}[i\Sigma^L_0(\omega)+i\Sigma^R_0(\omega)]$,
where $\Sigma^{L(R)}_{0}(\omega)=-i\Gamma^{L(R)}/2$ denotes the self-energy of
${\cal H}_0^{L(R)}$ for a flat wide band. We use $\Delta\equiv(\Gamma^L+\Gamma^R)/4$ as an energy unit.
The matrix reduction process corresponds to tracing out the reservoir degrees of freedom. However, the inter-reservoir coherence contained in the reservoir degrees of freedom remains in the coefficients $\beta_{pq}$ that appear at the $2\times 2$ corner blocks, e.g. $\beta_{14}=\xi^{L*}_{-}\xi^{R}_{+}$. The complete expressions for $\beta_{pq}$ are given in Ref.~\cite{hong11}.
Finally, the spectral function at the mediating atom is written as
\begin{equation}\label{eq8}
\rho_{m\uparrow}^{ss}(\omega)=(1/\pi){\rm Re}[({\bf M}^r)^{-1}]_{33},
\end{equation}
where ${\rm\bf M}^{r}$ is represented by the matrix ${\rm\bf M}_C$ with the addition of self-energy
$-i\Sigma_{pq}$ in each matrix element except $U_{j^\pm}^{L,R}$.
Then, the matrix ${\rm\bf M}^{r}$ consists of two $3\times 3$ blocks representing each single-reservoir Kondo system~\cite{hong10}. They share the central element representing the mediating Kondo atom.
Two $2\times 2$ blocks at the corners of ${\rm\bf M}^{r}$ establish the entanglement between two reservoirs. Hence, vanishing of the $2\times 2$ corner blocks corresponds to neglecting inter-reservoir coherence.

\begin{figure}
[b] \vspace*{4.3cm} \includegraphics{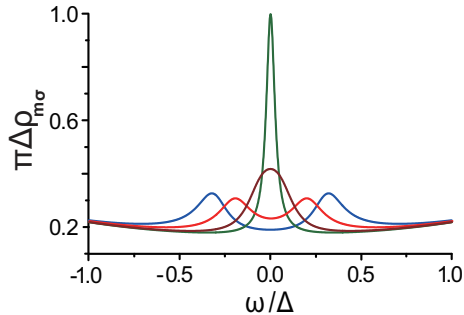} \vspace*{0.0cm} \caption{ (Color
online) Spectral functions of a non-EKS state. We adopt $\gamma_{_{LL(RR)}}=0.4$,
${\rm Re}U_{j^\pm}^{L,R}=1.5$, and ${\rm Re}[\beta_{pq}]=0.25$. The bias parameter is chosen as
$\gamma_{_{j}}=0$ (green), $\gamma_{_{j}}=0.1$ (brown), $\gamma_{_{j}}=0.2$ (red), and
$\gamma_{_{j}}=0.3$ (blue).
 }
\end{figure}

\section{Spectral functions}
Now, we study the EKS and non-EKS spectral functions using equation (\ref{eq8}). We first analyze the matrix ${\rm\bf M}^{r}$ at the atomic limit, i.e., ${\rm\bf M}_{C}$ with $U_{j^\pm}^{L,R}=U/4$ in equation (\ref{eq6}). For simplicity, we consider the symmetric case, $\gamma_{_{LL}}=\gamma_{_{RR}}$. Then, the inverse of ${\rm\bf M}_{C}$ yields five poles at $\omega'=0$, $\omega'=\pm\gamma_{_{LL}}$, and $\omega'\approx\pm U/2$ for large $U$. The first pole, i.e., the Kondo peak, has  a spectral weight~\cite{hong11}
\begin{equation}
Z=\left[1+\frac{U^2\{\gamma^2_{_{LL}}+\gamma_{_{RR}}^2+2(\gamma_{_{LR}}-\gamma_{_{j}})^2\}}
{8(\gamma_{_{LL}}\gamma_{_{RR}}+\gamma_{_{LR}}^2-\gamma_{_{j}}^2)^2}\right]^{-1},
\label{z0}
\end{equation}
which gives $Z=4\gamma_{_{LL}}^2/U^2+4\gamma_{_{LR}}^2/U^2$ at equilibrium ($\gamma_{_{j}}=0$) for large $U$. Interestingly, equation (\ref{z0}) gives $Z=4\gamma_{_{LL}}^2/U^2$ under bias ($\gamma_{_{j}}=\gamma_{_{LR}}$). This fact explicitly exhibits the disappearance of $4\gamma_{_{LR}}^2/U^2$, i.e., the contribution by entanglement, when a bias is applied. As a result, both Kondo peak weights of the EKS under bias and non-EKS have the same spectral weight, $Z=4\gamma_{_{LL}}^2/U^2$.

\begin{figure}
[t] \vspace*{5cm} \includegraphics{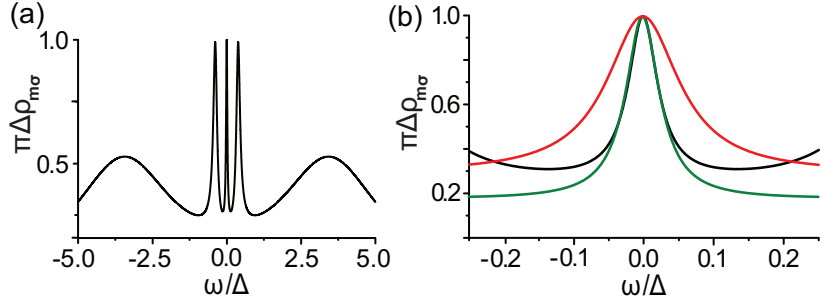} \vspace*{0.0cm} \caption{ (Color
online) (a) A typical five-peak spectral function of an EKS state. We set $\gamma_{_{j}}=\gamma_{_{LR}}=0.5$; other parameters are the same as in Fig.~3. (b) Comparison of the Kondo peaks for the EKS state at equilibrium (red),
EKS state under bias (black), and non-EKS state at equilibrium (green).}
\end{figure}

We first plot the non-EKS spectral functions exhibiting bias-dependent Kondo peak splitting. The bias dependence of the non-EKS state appears as the antidiagonal element $\gamma_{_{j}}$. Therefore, we adopt null $2\times 2$ corner blocks but have a finite $\gamma_{_{j}}$ reflecting bias effect. Then, we obtain bias-dependent Kondo peak splitting, as shown in Fig.~3. Next, we obtain the EKS spectral function under bias. A typical form of the EKS spectral function under bias is plotted in Fig.~4(a). Unidirectional motion of the EKS gives a bias-independent spectral function until a quasiparticle is excited by the bias. The five-peak spectral function shown in Fig.~4(a) is completely different from the bias-dependent split Kondo peak shown in Fig.~3, the previous studies~\cite{schoeller,wingreen2,fujii,han,anders}, and the three-terminal experiment~\cite{letur}. The essential difference is the appearance of two coherent side peaks near $\omega'=\pm\gamma_{_{LL}}$, which are crucial in explaining the experimental results for various mesoscopic Kondo systems. The two coherent side peaks signify the inter-reservoir coherence. In Fig.~4(b), we superimpose the three Kondo peaks: That of Fig.~4(a) (black), its equilibrium counterpart (red), and that of Fig.~3 for $\gamma_{_{j}}=0$ (green) for comparison. Figure~4(b) supports the disappearance of $4\gamma_{_{LR}}^2/U^2$ by bias, as discussed above.

\section{Conclusions}
We reveal that the methods used in the previous approaches are inappropriate for studying mesoscopic Kondo systems with inter-reservoir coherence [Fig.~1(c)]. These previous studies make use of the Hershfield model [Fig.~1(b)] in which non-EKS tunneling occurs and inter-reservoir coherence is not taken into account. Thus, obtaining a bias-dependent split Kondo peak in the spectral function is natural. We clarify that two additional coherent peaks appear in the spectral function as an effect of inter-reservoir coherence. We will obtain the $dI/dV$ line shapes of various mesoscopic Kondo systems and fit the experimental data in a separate study.

\ack The author thanks P. Coleman for suggesting the entangled Kondo singlet and
A. Millis, P. Kim, S.-W. Cheong, and P. Fulde for valuable discussions.
This research was supported by the Basic Science Research Program through the NRF, Korea
(2012R1A1A2005220), and was partially supported by a KIAS grant funded by MEST.

\end{document}